\documentclass[10pt,twoside]{article}
\begin{document}
\begin{center}
{\bf The Nobel Prize in Physics 2002 for the observation of
cosmic neutrinos}
\bigskip

B. Ananthanarayan\\
Centre for Theoretical Studies, Indian Institute of Science\\
Bangalore 560 012, India
\bigskip

\end{center}
\bigskip

\begin{abstract}
A brief description of the work for which the first
half of the Nobel prize for physics for the year 2002 is presented.
\end{abstract}

The first half of the Nobel Prize in physics for 2002 has been shared 
by Raymond Davis, Jr. and Masatoshi Koshiba for their observations of
cosmic neutrinos.  On two earlier occasions, Nobel prizes have
been awarded for discovery of neutrinos from controlled experiments:
in 1988 to Leon M. Lederman, Melvin Schwartz and Jack Steinberger for
the discovery of the muon type neutrino and the neutrino beam method, and
in 1995 to Frederick Reines for the discovery of the electron type
neutrino (discovered jointly with the by then deceased Clyde L. Cowan, Jr.),
who shared the prize with Martin Perl for his discovery of the tau lepton.
Neutrinos are particles that participate only in the weak
interactions, apart from participating in the gravitational
interactions.  The existence of the electron type neutrino was postuled 
in 1930 by Wolfgang Pauli in order to account for energy and angular momentum
conservation in nuclear beta decay.  They do not carry electric charge
and the so-called color charge of the strong interactions.  With the
discovery of the muon in the '40's and the tau lepton in the
'70's, heavier cousins of the electron, the electron type neutrino 
found cousins in the muon type and tau type neutrinos.

The name of Ray Davis is a legendary one and
is associated with the ``solar neutrinos'' which he observed
through decades of painstaking observations in a
deep underground experiment located in the Homestake Mine, in South
Dakota, USA.  As the name suggests, the neutrinos that he sought to
observe are those that are produced in copious numbers in the sun during
the many processes that power it.  An extraordinarily large number
of low-energy neutrinos are produced which are produced in the core and
travel outwards practically unhindered, due to the very small interaction
cross-section of these weakly inteacting particles.  In order to have 
observable events induced by this small cross-section, despite the large
flux of neutrinos at the distance of 1 astronomial unit
(distance between the earth and the sun), one must have a very 
large number of targets.  Davis'
achievement was to construct such an experiment:  the targets were the
nuclei of chlorine atoms present in a common cleaning liquid, 
tetrachloroethylene, which he stored in a large tank in the underground 
experiment, and the products of the neutrino capture by the nuclei of 
the chlorine atoms were a radioactive isotope of the
nobel gas argon.  The idea of
using chlorine nuclei as targets was due to the Italian born Soviet
physicist Bruno Pontecorvo and was realized ingeneously by the experiment
of Davis.   It must be noted, however, that the number of
reaction products in the over 615 tonnes of fluid that were exposed
to the solar neutrinos, during a run of a month was of the order of
20 argon atoms.  The ingenuity of Davis was to devise a technique to
flush out these argon atoms.  He used a method by which the tank was
flushed with helium, another noble gas.  He was able to demonstrate
the efficacy of this method to actually collect the argon atoms so
produced.  He thus established that solar neutrinos in fact exist, the
first experimental demonstration of the prediction from the models
of the sun.  The ``solar neutrino problem'' however, was that the
numbers of argon atoms so collected by Davis was systematically lower
than what was predicted by the models, by as much as a factor of
a third or a fourth.  Through dedicated experiments, he was able to 
show that none of the argon atoms produced were being lost during the 
procedure of extraction, and that the problem was to stay for many decades.

The existence of the solar neutrino problem was confirmed by the
experiment Kamiokande, located in the Mozumi mine 
in Kamioka-cho, Gifu, in Japan
(Kamiokande: Kamioka Nucleon Decay Experiment) and with which the person
sharing the prize with Davis, M. Koshiba is identified.
The experiment was established to search for nucleon decay, predicted
by grand unified theories of the fundamental interactions.
The experiment proved to be versatile enough to serve as a
``neutrino telescope.'' The experiment uses
the principle that neutrinos coming
from cosmic sources when entering a large tank of water would interact
with the electrons in the water and scatter elastically off the
electrons. These ultra-relativistic electrons emit characteristic
\v{C}erenkov radiation which is detected by photomultiplier tubes on the
periphery of the container of the water.  
At Kamiokande, 3000 tonnes of water were surrounded with 1000
photomultiplier tubes.  Unlike the Davis experiment,
the measurements here were ``real time'' and also had information of
the direction from which the observed neutrino was coming.  These
experiments were
able to spectacularly confirm the existence of the solar neutrino
problem.  Another unexpected and brilliant observation of the
Kamiokande experiment was that of 11 neutrino events in the year
1987 associated with the collapse of the neutrino-sphere of the
supernova 1987A.  The supernova which exploded in the Large
Magellenic cloud, one of the two companions of the Milky Way galaxy,
was also observed optically and in other regions of the electromagnetic
spectrum, and its light curve was carefully monitored. This was the
first ever supernova to be explored in this remarkable manner.
The standard picture of the supernova explosion requires that a significant
fraction of the energy transport out of the supernova be in the
form of that transported by neutrinos in a spectacular burst.
The event rates calculated for Kamiokande proved to be
in agreement with what was observed.  Supernova neutrinos were also
observed by the experiments IMB (Irvine-Michigan-Brookhaven) and
the Baksan mine experiments, which confirmed that the neutrinos
observed by Kamiokande were indeed supernova neutrinos.

The direct observation of the neutrino oscillations by the
Sudbury Neutrino Observatory has recently resolved the solar neutrino
problem in favor of a particle physics scenario whereby the electron
type neutrino that is produced at the core of the sun has a finite
probability of oscillating into a muon or tau type neutrino. 
Furthermore, no modification of the standard solar model based on
which the fluxes of neutrinos are calculated is necessary (see article in
Resonance, Vol. 7, No. 10, pp. 79 (2002)).

The award of the Nobel prize to Davis is one that would do the
prize proud.  During the 30 or years of the running
of his experiment, around 2000 argon atoms were collected! At age 87, 
he stands as an example of the rare dedication and
perseverence that is required to make a
truly outstanding contribution to science.  A chemist by training, 
Davis is Professor Emeritus at the University of Pennsylvania.

The award to Koshiba is no less deserving;
the courage to design outstanding and brilliant experiments to search
for the unusual and to prove the versatility of experiments.
At the age of 76, Koshiba is Professor Emeritus at the University of Tokyo.

\bigskip

\bigskip

\noindent For recent articles in Resonance of related interested, see

\begin{enumerate}

\item B. Ananthanarayan and Ritesh K. Singh, Vol. 7, No. 10,
pp. 79 (2002).

\item S. M. Chitre, Resonance, Vol. 7, No. 8, pp. 67 (2002).

\end{enumerate}
\end{document}